\documentclass[sigconf]{acmart}

\usepackage{tabularx}
\usepackage{array}
\usepackage{enumitem}
\usepackage{hyperref}

\AtBeginDocument{%
  }

\setcopyright{cc}
\setcctype{by-nd}
\copyrightyear{2025}
\acmYear{2025}
\acmConference[CHI '25]{CHI Conference on Human Factors in Computing Systems}{April 26-May 1, 2025}{Yokohama, Japan}
\acmBooktitle{CHI Conference on Human Factors in Computing Systems (CHI '25), April 26-May 1, 2025, Yokohama, Japan}
\acmDOI{10.1145/3706598.3713782}
\acmISBN{979-8-4007-1394-1/25/04}

\begin{document}

\title[Designing Health Technologies for Immigrant Communities]{Designing Health Technologies for Immigrant Communities: Exploring Healthcare Providers’ Communication Strategies with Patients}

\author{Zhanming Chen}
\email{chen8475@umn.edu}
\orcid{0000-0002-9913-7239}
\affiliation{%
  \institution{University of Minnesota}
  \city{Minneapolis}
  \state{MN}
  \country{USA}
}

\author{Alisha Ghaju}
\email{ghaju001@umn.edu}
\orcid{0009-0003-0870-4088}
\affiliation{%
  \institution{University of Minnesota}
  \city{Minneapolis}
  \state{MN}
  \country{USA}
}

\author{May Hang}
\email{may.hang@hennepin.us}
\orcid{0009-0006-8809-679X}
\affiliation{%
  \institution{NorthPoint Health and Wellness Center}
  \city{Minneapolis}
  \state{MN}
  \country{USA}
}

\author{Juan F. Maestre}
\email{j.f.maestreavila@swansea.ac.uk}
\orcid{0000-0002-5403-9387}
\affiliation{%
  \institution{Swansea University}
  \city{Swansea}
  \country{UK}
}

\author{Ji Youn Shin}
\email{shinjy@umn.edu}
\orcid{0000-0003-4978-3897}
\affiliation{%
  \institution{University of Minnesota}
  \city{Minneapolis}
  \state{MN}
  \country{USA}
}

\renewcommand{\shortauthors}{Chen et al.}

\begin{abstract}
  Patient-provider communication is an important aspect of successful healthcare, as it can directly lead to positive health outcomes. Previous studies examined factors that facilitate communication between healthcare providers and patients in socially marginalized communities, especially developing countries, and applied identified factors to technology development. However, there is limited understanding of how providers work with patients from immigrant populations in a developed country. By conducting semi-structured interviews with 15 providers working with patients from an immigrant community with unique cultural characteristics, we identified providers’ effective communication strategies, including acknowledgment, community involvement, gradual care, and adaptive communication practices (i.e., adjusting the communication style). Based on our findings, we highlight cultural competence and discuss design implications for technologies to support health communication in immigrant communities. Our suggestions propose approaches for HCI researchers to identify practical, contextualized cultural competence for their health technology design.
\end{abstract}




\maketitle

\section{INTRODUCTION}

Patient-provider communication is one of the most important factors determining the quality of healthcare~\cite{ding_etal_BoundaryNegotiationPatientProvider_2019}. It is a collaborative decision-making process in which both patients and providers contribute their medical expertise and exchange information~\cite{jenerette_mayer_PatientProviderCommunicationRise_2016}. Patients’ health outcomes, feelings, and satisfaction with health management are strongly influenced by their interaction with providers~\cite{capone_PatientCommunicationSelfefficacy_2016}.

Immigrant communities often face difficulties communicating with healthcare providers in their new environment. Currently, there are more than 46 million immigrants in the U.S.~\cite{americanimmigrationcouncil_TakeLookHow_2022}. The percentage of immigrants in the total U.S. population (13.8\%) has nearly tripled since 1970, when it was 4.7\%~\cite{moslimani_passel_WhatDataSays_2024}. Immigrant communities face several healthcare challenges: 15\% of all immigrants are uninsured; among those with limited English proficiency (LEP), 21\% lack insurance; and 24\% of low-income immigrants lack access to healthcare services~\cite{pillai_etal_HealthHealthCare_2023}. Limited access to health insurance, language barriers, and low socioeconomic status contribute to negative healthcare experiences and mistrust of providers~\cite{khullar_chokshi_ChallengesImmigrantHealth_2019, phillips_etal_PatientproviderCommunicationPatterns_2021}. Additionally, less effective patient-provider communication can lead to negative healthcare outcomes, because many health conditions, including cancer, require long-term care and consistent communication with providers~\cite{singh_etal_ImmigrantHealthInequalities_2013}. Therefore, the healthcare experiences of immigrant communities—particularly their communication with healthcare providers—are a significant public health issue~\cite{lebron_etal_ImmigrationImmigrantPolicies_2023}. 

To address these obstacles, studies often suggested approaches for alleviating information access barriers~\cite{chandwani_kumar_StitchingInfrastructuresFacilitate_2018, talhouk_etal_SyrianRefugeesDigital_2016} and training healthcare providers in improved communication skills~\cite{kron_etal_UsingComputerSimulation_2017, muke_etal_DigitalTrainingNonSpecialist_2020}. Others have focused on providing additional resources, such as recruiting bilingual providers to facilitate communication between immigrant patients and their providers, as well as offering medical content in different languages~\cite{anderson_etal_CulturallyCompetentHealthcare_2003, phillips_etal_PatientproviderCommunicationPatterns_2021, thonon_etal_ElectronicToolsBridge_2021}.

HCI studies have identified barriers that negatively influence patients’ communication experiences with their providers: mistrust~\cite{burgess_etal_CareFrictionsCritical_2022}, health literacy~\cite{park_etal_HealthLiteracySupporting_2017}, and cultural differences~\cite{phillips_etal_PatientproviderCommunicationPatterns_2021}. In response, researchers have found that the use of electronic health records (EHRs)~\cite{street_etal_ProviderInteractionElectronic_2014}, agent-based virtual health education~\cite{bickmore_etal_TakingTimeCare_2009, sakpal_VirtualPatientsTeach_2012}, instant messaging applications~\cite{ding_etal_BoundaryNegotiationPatientProvider_2019}, and medical documents interpretation tools~\cite{hong_etal_SupportingFamiliesReviewing_2017} are effective solutions for facilitating communication between patients and providers. Moreover, in recent years, researchers have started considering patients’ stories and values that need to be captured and effectively utilized during their care process~\cite{berry_etal_SupportingCommunicationValues_2019, hong_etal_UsingDiariesProbe_2020}. This is particularly true for vulnerable populations, such as cultural minority groups and pediatric patients with chronic conditions. HCI researchers have shifted their focus from the effective delivery of medical information and education to capturing more contextualized health information about patients' everyday lives, such as their values, hobbies, and religious practices~\cite{seo_etal_ChallengesParentChildCommunication_2021, smith_etal_WhatSpiritualSupport_2021}. 

Cultural competence is a set of attitudes, behaviors, and interventions that enable individuals to interact with people with different cultural backgrounds~\cite{anderson_etal_CulturallyCompetentHealthcare_2003}. This approach is especially relevant in developed countries, such as the U.S., where many culturally marginalized populations (e.g., immigrants) live alongside those from mainstream cultures~\cite{balcazar_etal_CulturalCompetenceDevelopment_2009}. Medical literature has examined the factors that increase cultural competence, including cultural awareness, knowledge, and skills/behaviors~\cite{alizadeh_chavan_CulturalCompetenceDimensions_2016, shen_CulturalCompetenceModels_2015}. Although HCI studies did not explicitly mention the term “cultural competence” in healthcare, studies examined factors that influence health management experiences in vulnerable populations and technology-mediated approaches to address challenges in those contexts (e.g., the Global South~\cite{ismail_etal_BridgingDisconnectedKnowledges_2018, verdezoto_etal_InvisibleWorkMaintenance_2021}, low SES~\cite{grimes_etal_EatWellSharingNutritionrelated_2008, harrington_etal_EngagingLowIncomeAfrican_2019}). To alleviate the identified barriers, researchers have developed various approaches to improve healthcare experiences for these populations, including patient education materials~\cite{kumar_etal_ProjectingHealthCommunityled_2015}, enhanced access to care~\cite{tiwari_sorathia_VisualisingSystematizingPoor_2014, yadav_etal_ShouldVisitClinic_2022}, and communication support tools for both patients and providers~\cite{okeke_etal_IncludingVoiceCare_2019}. Nevertheless, limited studies have examined healthcare providers’ coping strategies in communicating with their patients in socially marginalized communities. Given the barriers immigrant patients face in managing their health, understanding the unique aspects of patient-provider communication in immigrant communities is necessary to propose design implications for technologies.

This study aims to explore healthcare providers' communication strategies and implications for health technology design. To gain a comprehensive understanding of participants' perspectives and insights, we adopted a constructivist approach. We conducted 15 semi-structured interviews with healthcare providers who care for immigrant patients with unique cultural characteristics. Through these interviews, we identified key communication practices used by providers to enhance cultural competence in caring for immigrant communities, including addressing patients' fears by highlighting commonalities between providers' and patients' healthcare practices, and utilizing storytelling as a tool to help patients articulate and understand their symptoms. The study is guided by the following three main research questions: 
 
\begin{itemize}
  \item What are the healthcare providers’ communication strategies to better engage their immigrant patients? (RQ1)
  \item How might we incorporate the concept of cultural competence in health technologies designed for immigrant communities? (RQ2)
  \item What are the design implications of health technologies facilitating communication between providers and their immigrant patients? (RQ3)
\end{itemize}

By answering these research questions, our contribution is threefold. First, we present an empirical understanding of healthcare providers’ communicating strategies developed from their experiences with immigrant patients. Second, we suggest approaches for HCI researchers to incorporate cultural competence in health technology design, particularly for socially marginalized communities. Third, we identify critical design implications towards culturally competent health communication technology. In the following sections, we present related work, data collection methods and analysis, followed by our main findings and discussion.

\section{RELATED WORK}

We reviewed HCI and medical studies that 1) examined design implications and technology-mediated solutions to support improved patient-provider communication, and 2) explored health management in socially marginalized populations, and technologies to improve their current practices.

\subsection{Technologies to Support Patient-Provider Communication}

Patient-provider communication is a collaborative process between patients and providers, with providers contributing medical expertise and knowledge, while patients contribute knowledge of their preferences, needs, and values (e.g., health and wellness experiences, day-to-day routines, and lifestyles) to guide their healthcare decisions~\cite{ballegaard_etal_HealthcareEverydayLife_2008, jenerette_mayer_PatientProviderCommunicationRise_2016, schroeder_etal_SupportingPatientProviderCollaboration_2017}. Patient-provider communication has significant impacts on patients’ health outcomes~\cite{lim_etal_ItJustSeems_2016}. Improved communication between patients and providers is positively associated with patients’ adherence to treatments~\cite{jenerette_mayer_PatientProviderCommunicationRise_2016}, patients’ positive feelings and attitudes~\cite{blackwell_etal_UsingTextMessaging_2020}, and higher level of patient satisfaction in their care~\cite{carcone_etal_EffectivePatientProvider_2016}. It could also contribute to overall positive healthcare outcomes~\cite{henry_etal_AssociationNonverbalCommunication_2012}. Negative patient-provider communication can cause medical errors, and poor healthcare outcomes~\cite{saha_etal_PatientPhysicianRelationships_2003}. With an improved understanding of patient-provider communication and its significance in patients’ health outcomes, researchers in HCI and related fields have shifted their focus from traditional care where providers focus on patients’ symptoms and diagnosis, to patient-centered collaborative models~\cite{ jenerette_mayer_PatientProviderCommunicationRise_2016,vranceanu_etal_IntegratingPatientValues_2009}. Their research examined factors that influence patient-provider communication (e.g., patients’ socio-cultural characteristics (e.g., religion)~\cite{burgess_etal_CareFrictionsCritical_2022, phillips_etal_PatientproviderCommunicationPatterns_2021}, different expectations of patients and providers~\cite{park_etal_HealthLiteracySupporting_2017}, health literacy~\cite{hironaka_paasche-orlow_ImplicationsHealthLiteracy_2008, pollack_etal_ClosingGapSupporting_2016}, and privacy~\cite{okeke_etal_IncludingVoiceCare_2019}, and technologies to support actors involved in care process, including patients, caregivers, and providers. These studies can be categorized into two groups: 1) health information exchange between patients and providers and 2) providers’ comprehension of patients’ value in different populations.

The first group of studies focuses on ways to promote patients’ active health information tracking and sharing during communication with their providers. These studies pointed out health literacy~\cite{park_etal_HealthLiteracySupporting_2017}, cultural difference~\cite{schouten_meeuwesen_CulturalDifferencesMedical_2006, cooper_etal_AssociationsCliniciansImplicit_2012}, mistrust~\cite{burgess_etal_CareFrictionsCritical_2022, schroeder_etal_SupportingPatientProviderCollaboration_2017}, timing~\cite{fu_etal_UnveilingHighspeedFollowConsultation_2022}, and other factors as key components that result in less effective health information communication between patients and providers. Researchers explored patients and providers’ experiences in different contexts, including pediatric care~\cite{hong_etal_SupportingFamiliesReviewing_2017, kong_karahalios_AddressingCognitiveEmotional_2020,seo_etal_LearningHealthcareProviders_2021}, emergency visits~\cite{park_etal_HealthLiteracySupporting_2017}, and multiple chronic diseases care~\cite{berry_etal_SupportingCommunicationValues_2019, fu_etal_UnveilingHighspeedFollowConsultation_2022}, and suggested technology-mediated solutions. This line of research showed the effectiveness of designing EHRs~\cite{street_etal_ProviderInteractionElectronic_2014}, patient portals~\cite{sun_etal_MessagingYourDoctors_2013}, agent-based virtual health education~\cite{bickmore_etal_TakingTimeCare_2009, sakpal_VirtualPatientsTeach_2012}, instant messaging applications~\cite{ding_etal_BoundaryNegotiationPatientProvider_2019}, digital journals~\cite{chung_SupportingPatientproviderCommunication_2017, hong_etal_UsingDiariesProbe_2020, schroeder_etal_SupportingPatientProviderCollaboration_2017}, and medical documents interpretation tools~\cite{hong_etal_SupportingFamiliesReviewing_2017}. For instance, Seo et al. identified providers’ effective strategies when communicating with pediatric patients by proactively asking patients’ conditions, learning about patients’ communication preferences, and sharing patients’ information with clinic team members~\cite{seo_etal_LearningHealthcareProviders_2021}. This study suggested the importance of providers’ strategies to understand child patients’ initial responses and learn patients’ communication preferences, to adjust their communication styles. Another study showed increased levels of assertiveness implemented in health information technology as benefits that influence providers’ satisfaction and reduce medical errors~\cite{calisto_etal_AssertivenessbasedAgentCommunication_2023}. Similarly, Schroeder et al. applied collaborative data review to increase the validity of the patient-generated data, thereby enhancing patients’ autonomy in the healthcare process~\cite{schroeder_etal_SupportingPatientProviderCollaboration_2017}. Others also developed a streaming video-based visualization tool that facilitated communication between parents and providers about children’s developmental delays through the annotation of children’s gestures and verbalization in the clinical visits’ videos~\cite{kong_karahalios_AddressingCognitiveEmotional_2020}. 

In addition to facilitating medical information sharing, more recent studies have started to examine approaches incorporating patients’ values in the care process. Studies showed that understanding patients’ non-medical information, including their values, critical life events, personalities, and information on their family members, is an effective approach to long-term chronic illness care where collaborative practices between patients and providers are necessary~\cite{berry_etal_SupportingCommunicationValues_2019, lim_etal_ItJustSeems_2016, seo_etal_ChallengesParentChildCommunication_2021, vranceanu_etal_IntegratingPatientValues_2009}. This body of research has suggested the visualization of patients' values~\cite{berry_etal_SupportingCommunicationValues_2019, ryu_etal_YouCanSee_2023}, social signal recognition and analysis tools~\cite{bedmutha_etal_ConverSenseAutomatedApproach_2024, faucett_etal_ShouldListenMore_2017, patel_etal_VisualFeedbackNonverbal_2013}, and collaboration with providers to share information about patients' values~\cite{seo_etal_LearningHealthcareProviders_2021} as useful technology-mediated solutions. For example, Berry et al. advocated for providers to use surveys, cameras, and wearable devices to collect information about patients' value, which is often not disclosed because patients do not feel it directly relates to their symptoms~\cite{berry_etal_SupportingCommunicationValues_2019}. Similarly, Ryu et al. centered values of patients with multiple chronic health conditions throughout their self-care practices and health status, and suggested that providers facilitate understanding of patients’ values through visualization technology to assist patients in choosing a focus from their issues as a priority~\cite{ryu_etal_YouCanSee_2023}. 

\subsection{Technologies to Support Healthcare Management in Marginalized Communities}

The individuals’ socio-cultural background significantly influences health inequalities~\cite{centersfordiseasecontrolandprevention_SocialDeterminantsHealth_2024, martin-hammond_purnell_BridgingCommunityHistory_2022}. Studies in HCI and related fields have examined how these characteristics (e.g., socioeconomic status, education, physical environment, ethnicity, internet access, social support networks) affect their health management experiences~\cite{braveman_etal_SocialDeterminantsHealth_2011}, and suggested technology-mediated solutions to support effective health outcomes. These studies explored healthcare experiences in various marginalized communities, such as immigrants~\cite{fernandez-gutierrez_etal_EffectMHealthIntervention_2019, ghaju_etal_SupportingHealthcareProviders_2024, morales_zhou_HealthPracticesImmigrant_2015, tachtler_etal_UnaccompaniedMigrantYouth_2021, talhouk_etal_SyrianRefugeesDigital_2016}, residents in the Global South~\cite{ismail_etal_BridgingDisconnectedKnowledges_2018, verdezoto_etal_InvisibleWorkMaintenance_2021}, people of color~\cite{harrington_etal_EngagingLowIncomeAfrican_2019}, individuals with LEP~\cite{bickmore_etal_TakingTimeCare_2009, kim_etal_ItsMyLanguage_2024}, women~\cite{blackwell_etal_UsingTextMessaging_2020, guendelman_etal_ListeningCommunitiesMixedMethod_2017}, and older adults~\cite{latulipe_etal_DesignConsiderationsPatient_2015, lee_etal_StepsParticipatoryDesign_2017}. 

These studies investigated various factors that impact individuals from these marginalized communities. These factors include health literacy~\cite{carballo_mboup_InternationalMigrationHealth_2005}, limited access to health insurance~\cite{mohammadi_etal_HealthCareNeeds_2016}, cultural beliefs~\cite{harrington_etal_ItsKindCodeSwitching_2022}, cultural competence~\cite{sakpal_VirtualPatientsTeach_2012, truong_etal_InterventionsImproveCultural_2014}, and language~\cite{ismail_kumar_EngagingSolidarityData_2018} as unique challenges each community faced. For example, researchers examined patients’ unique cultural practices in managing health (e.g., alternative medicines, spiritual practices) in the Global South as critical considerations for achieving satisfactory care~\cite{ismail_kumar_EngagingSolidarityData_2018}. Through design workshops, Harrington et al. identified factors that affect healthcare experiences in the African-American older adult population~\cite{harrington_etal_EngagingLowIncomeAfrican_2019}.  

Research in this area has also explored different technology-mediated approaches and practical suggestions to alleviate challenges and propose an improved way of managing health in each marginalized context. The goals of these interventions include educating patients on health~\cite{fernandez-gutierrez_etal_EffectMHealthIntervention_2019, kumar_etal_ProjectingHealthCommunityled_2015, molapo_etal_DesigningCommunityHealth_2016}, improving patients’ medical access~\cite{tiwari_sorathia_VisualisingSystematizingPoor_2014, yadav_etal_ShouldVisitClinic_2022}, communication support tools for both patients and providers~\cite{okeke_etal_IncludingVoiceCare_2019, patel_etal_VisualFeedbackNonverbal_2013}, and training contents to improve providers’ communication skills~\cite{kron_etal_UsingComputerSimulation_2017, muke_etal_DigitalTrainingNonSpecialist_2020}. For example, Kumar et al. used videos to distribute medical information and education on topics such as breastfeeding and birth preparedness for pregnant women in the Global South~\cite{kumar_etal_ProjectingHealthCommunityled_2015}. Another study suggested improved ways to support women with low literacy by describing symptoms with photos and providing audio-based medical advice from providers~\cite{tiwari_sorathia_VisualisingSystematizingPoor_2014}. 

Studies on patient-provider communication discussed effective strategies for engaging patients in their care and identified unique needs and characteristics of each population. However, there is limited understanding of how healthcare providers and immigrant patients in a developed country effectively communicate. Also, only a few studies in HCI have discussed healthcare providers’ coping strategies for communicating with their patients in socially marginalized communities. Although several HCI studies have examined approaches to enhance cultural competence in different contexts, such as assisting with students' wellness~\cite{gaither_etal_RoleCulturalCompetency_2018} and facilitating awareness of refugees' experiences~\cite{faltaous_etal_VirtualRealityCultural_2018}, it has not been thoroughly examined in the context of health communication. Given the vast array of barriers and challenges immigrant patients face in communicating with providers, there is a need to further understand the characteristics of patient-provider communication in immigrant communities to suggest design implications for supportive technologies.

Building on studies that emphasized the potential for adopting technology to facilitate patient-healthcare provider communication and health management among underserved populations, we report our findings showing the experiences of healthcare providers working with immigrant communities. These experiences and strategies can further inform design implications for the design of supportive patient-provider communication technologies in culturally marginalized communities.

\section{METHODS}

\subsection{Study Context}

In this study, we interviewed 15 healthcare providers working with patients from the Hmong community, an ethnic group with origins in Southeast Asia, specifically Laos, Vietnam, and Thailand. During the Vietnam War, the Hmong supported American troops, and after the war, they were granted refugee status and migrated to the U.S. in large numbers. According to the U.S. Census Bureau, there were more than 300,000 Hmong Americans in 2022~\cite{uscensusbureau_AmericanCommunitySurvey_}. The Hmong community is different from most other American immigrant populations in that they have strongly retained their original sociocultural characteristics and experienced only minimal assimilation~\cite{guan_etal_AsianAmericanEnclaves_2023}. Nowadays, the vast majority of Hmong-Americans are agricultural workers, living substantially below the poverty line, and speaking little to no English~\cite{uscensusbureau_AmericanCommunitySurvey_}.

In health management, Hmong Americans experienced drastic changes and challenges regarding adapting their cultural practices and lifestyles to healthcare practice in the U.S.~\cite{johnson_HmongHealthBeliefs_2002}. For example, the Hmong community's adherence to their traditional health practices (e.g., herbal medication, religious practices, such as shamanism)~\cite{lee_vang_BarriersCancerScreening_2010, thorburn_etal_MedicalMistrustDiscrimination_2012}) prevents them from having necessary clinic visits and results in medication nonadherence. While many chronic diseases are treatable with early diagnosis and appropriate treatment, the health outcomes for Hmong Americans are often exacerbated by health inequities, including lack of medical insurance coverage, LEP, low education levels, and unequal access to treatments~\cite{lee_vang_BarriersCancerScreening_2010}. As a result, they face higher health risks compared to other immigrant populations, with elevated incidence and mortality rates for chronic conditions (e.g., diabetes, hypertension) and cancers\cite{brown_etal_DigitalDiabetesStorytelling_2023, lor_etal_WesternTraditionalHealers_2017}. Despite these challenges, the Hmong community also has many cultural strengths, such as social support, strong family relationships, and respect for elders and their wisdom. These strengths allowed the community to settle down in the new area and thrive with their community~\cite{tatman_HmongHistoryCulture_2004}. We selected the Hmong community in and around a major U.S. metropolitan area as our study context. This community offers unique perspectives on health management that are deeply rooted in their culture.

\subsection{Data Collection}

From Fall 2023 to Spring 2024, we conducted one-on-one semi-structured interviews to collect data that would allow us to understand healthcare providers’ communication practices with Hmong patients. The interview questions aimed to explore the healthcare providers’ lived experience with the Hmong immigrant community, their communication strategies to facilitate Hmong patients’ health management, and challenges and opportunities for improving patient care. Examples of the interview questions include: “What were your experiences when working with Hmong patients and caregivers?”, “What kinds of health technologies do you use when communicating with Hmong patients?” and “What are Hmong patients’ major barriers, difficulties, and concerns when managing their health?” To better understand the patients' clinic visit and their experiences, we also visited two local clinics where healthcare providers recruited for this study work. We followed a provider and observed the patient journey in the clinic (check-in at the front desk, going to the treatment room, having a consultation, performing an examination, getting a prescription, and following up with an appointment). This observation delivered an overview of immigrant patients’ clinic visits. Previous studies on immigrant patients' health management and patient-provider communication informed the interview questions (e.g.,~\cite{lee_vang_BarriersCancerScreening_2010, thorburn_etal_MedicalMistrustDiscrimination_2012, jenerette_mayer_PatientProviderCommunicationRise_2016}). Each interview session lasted between 45 and 90 minutes and was audio recorded for transcription. Our university’s Institutional Review Board granted ethical approval for this study.

\subsection{Participants}

We recruited 15 healthcare providers who work with Hmong patients in Minneapolis, Minnesota, in the U.S. (see Table \ref{tab:demog}). The eligibility criteria for study participation included being a healthcare provider for Hmong patients, being able to communicate in English, and willing to participate in an interview. We recruited participants from a local clinic and a hospital through the referral of our collaborators. These institutions specifically serve local underserved communities, including many Hmong and other immigrant patients. Although we did not intentionally recruit healthcare providers with the same cultural background as the Hmong patients, 10 out of 15 participants reported speaking English and Hmong as their primary languages and having strong ties to the community, being second- or third-generation immigrants. Once potential participants showed interest, we emailed and presented them with an overview of the study. After participants agreed and consented to participate in the study, we scheduled the Zoom interviews. We conducted interviews until data saturation was reached and no new or additional information was collected~\cite{charmaz_ConstructingGroundedTheory_2014}.

The participants included physicians, nurses, pharmacists, care coordinators, social workers, and interpreters. As we conducted the initial interviews, we learned about the importance of having different perspectives in interacting with patients. This led us to recruit a wide range of providers to capture a more holistic understanding of providers’ communication experiences with their patients. As a result, we recruited providers from a variety of roles: six physicians, two nurse practitioners, two registered nurses, two clinical interpreters, one pharmacist, two care coordinators, and one social worker (Provider 15 is counted twice due to her multiple roles). Among them, five worked in family medicine, three in obstetrics and gynecology, one in pediatrics, and one in psychiatry. Study participants had a median age of 42 years (range: 27–68 years), with the majority identifying as female (13/15) and Asian (12/15). The most common duration of practice with Hmong patients was 5–10 years (4/15). Their annual household incomes ranged from \$25,000 - \$35,000 (1/15), \$50,000 - \$75,000 (1/15), \$75,000 - \$99,000(1/15), \$100,000 - \$200,000 (5/15) to more than \$200,000 (3/15). Additionally, more than half (8/15) were married or in a domestic partnership, and nearly all (13/15) had at least a 4-year college degree.

\begin{table*}
  \caption{Demographic information of participants}
  \label{tab:demog}
  \setlength\tabcolsep{2pt}
  \begin{tabular*}{\linewidth}{@{\extracolsep{\fill}} llllllll }
    \toprule
    Participants & Occupation & Age & Gender & \vtop{\hbox{\strut Ethnicity/}\hbox{\strut Race}} & Language Spoken & \vtop{\hbox{\strut Years in Practices}\hbox{\strut with Hmong Patients}}  \\
    \midrule

    P1 & Physician & 53 & M & Asian & English & 20-30 years\\
    P2 & Registered Nurse & 31 & F & Asian & Hmong and English & 1-5 years\\
    P3 & Physician & 35 & F & White & English & 5-10 years\\
    P4 & Physician & 35 & M & Asian & Hmong and English & 10-20 years\\
    P5 & Care Coordinator & 46 & F & Asian & Hmong and English & 5-10 years\\
    P6 & Physician & 56 & F & White & English & 20-30 years\\
    P7 & Registered Nurse & 42 & F & Asian & Hmong and English & 10-20 years\\
    P8 & Nurse Practitioner & 35 & F & Asian & Hmong & 5-10 year\\
    P9 & Physician & 35 & F & Asian & English and Mandarin & 1-5 years\\
    P10 & Social Worker & 44 & F & White & English & 10-20 Years\\
    P11 & Pharmacist & 33 & F & Asian & Hmong and English & 1-5 Years\\
    P12 & Physician & 43 & M & Asian & Hmong and English & 5-10 years\\
    P13 & Nurse Practitioner & 27 & F & Asian & Hmong and English & 1-5 Years\\
    P14 & Clinic Interpreter & 64 & F & Asian & Hmong Mostly & 30+ years\\
    P15 & \vtop{\hbox{\strut Care Coordinator/}\hbox{\strut Interpreter}} & 68 & F & Asian & Hmong and English & 30+ years\\

  \bottomrule
  
\end{tabular*}
\bigskip
\emph{Note:} F=Female, M=Male.

\end{table*}

\subsection{Data Analysis}

Using a grounded theory approach~\cite{charmaz_ConstructingGroundedTheory_2014}, we conducted a thematic analysis to identify emerging themes from the interviews~\cite{braun_clarke_UsingThematicAnalysis_2006}. Our primary goal was to explore the meanings and experiences that healthcare providers developed through their interactions with Hmong patients. The constructivist approach allowed the research team the flexibility to understand how healthcare providers developed strategies, free from predetermined perspectives or biases that might influence the researchers' interpretations. The first and second co-authors conducted open coding on the transcripts to identify an initial set of codes. We then used NVivo software to assign these initial codes to relevant segments of data, employing a descriptive coding method. The initial set contained approximately 2,000 low-level codes, including “Herbal medication is an alternative to western medications,” “It is important for providers to use images with the Hmong patient, as it helps in the visualization of conditions,” and “Being able to share a language and a culture with patients helps providers in building a comfortable environment.” As we repeatedly coded the data, we followed the thematic analysis approach to identify patterns across the transcripts from different participants. We iteratively compared and contrasted each pair of codes to see whether they were mutually exclusive. By continuously aggregating codes, we were able to reduce our original 2000 low-level codes to 280 mid-level codes. During three months of iterative coding, we regularly met and compared generated codes to resolve any discrepancies (e.g., interpreting participants’ words in different ways). Finally, we performed affinity clustering to identify emerging categories and themes. We defined and named high-level themes, such as “Understanding the value of strong social ties” and “Adapting communication style based on the characteristics of patients' language.”

\section{FINDINGS}

We identified four communication strategies that facilitate providers’ diagnosis, treatment, and care for the Hmong community. Our findings showed that providers actively developed and tried different tactics to engage their patients into the care process, which includes: 1) acknowledging for patients’ fears about Western medicine and being flexible with patients’ cultural practices; 2) understanding the value of
strong social ties and collective decision-making practices in the community; 3) establishing trustworthiness through gradual and inclusive long-term care; 4) adapting communication style based on the particular characteristics of the Hmong language. See Table \ref{tab:strate} for a summary of the providers' strategies and their specific communication practices. We first introduce the challenges providers face with Hmong patients, followed by providers’ stories on how they developed helpful communication strategies, illustrated by quotes.


\begin{table*}
  \caption{Healthcare providers’ strategies}
  \label{tab:strate}
  \begin{tabular}{p{6cm}p{11cm}}
    \toprule
    Healthcare Providers’ Strategies & Communication Practices \\
    \midrule

  
  \textbf{Strategy 1.} Acknowledging for patients’ fears about Western medicine and being flexible with patients’ cultural practices  
  & 
  \begin{itemize}
      \item Communicating behavioral guidelines at practical levels by contextualizing instructions based on norms of the community
      \item Showing understanding of Hmong patients’ cultural beliefs and values
      \item Reducing fears by explaining common factors between patients' and providers' practices
      \item Providing further information on the origin of cultural practices
  \end{itemize} \\ \hline
  \textbf{Strategy 2.} Understanding the value of strong social ties and collective decision-making practices in the community 
  &
  \begin{itemize}
      \item Actively involving a community or family leader 
      \item Sharing anecdotes from the community to resonate with patients
      \item Increasing relatedness using providers’ connection with the community
      \item Constantly increasing understanding of unique cultural practices
  \end{itemize} \\ \hline
  \textbf{Strategy 3.} Establishing trustworthiness through gradual and inclusive long-term care 
  &
  \begin{itemize}
      \item Gradually introducing new medications and therapies
      \item Offering a time frame to consider possible treatment options
      \item Communicating in a respectful manner
      \item Frequently following up via phone calls
  \end{itemize} \\ \hline
  \textbf{Strategy 4.} Adapting communication style based on the particular characteristics of the Hmong language 
  &
  \begin{itemize}
      \item Using storytelling to describe symptoms
      \item Capturing patients’ non-verbal cues
      \item Adjusting tone when discussing diagnoses
      \item Using visual materials to explain medical information
  \end{itemize} \\ 
    
  \bottomrule
  
\end{tabular}
\end{table*}

\subsection{Strategy 1: Acknowledging for Patients’ Fears about Western Medicine and Being Flexible with Patients’ Cultural Practices}

According to the providers, many Hmong patients are reluctant to follow prescriptions and advice suggested by their providers — lab tests and specialist visits — due to unfamiliarity with, and fear of, Western medicine and lack of health-related education, whereas have positive attitudes towards cultural therapies, such as herbal remedies and shamanism (e.g., healing ceremonies) passed down by their ancestors. Patients’ preferences towards cultural practices and non-compliance with prescriptions are more challenging, where long-term treatment is required for effective outcomes for symptoms, such as chronic illness care. For example, with diabetes and cardiovascular diseases, lifelong follow-up care is critical; regular exercise, diet, and medication intake are required to maintain healthy lives. Despite the needs, patients tend to wait until they experience notable symptoms and do not follow providers’ suggestions, opting for traditional herbal remedies as their first choice. 

Providers mentioned that, because many illnesses do not have detectable symptoms in their early stages, patients decide to reduce or even stop their medication, without discussing it with their providers. Eventually, the illness progresses to the point where it causes serious complications that are not easily treated. From their experiences with many Hmong patients, providers learn the benefits of acknowledging patients’ beliefs in cultural practices, and possible fears and unfamiliarity with Western medicine. This led providers to develop different practices that help them engage their patients in care: communicating behavioral guidelines at practical levels by contextualizing their instructions based on norms of the Hmong community, showing understanding of Hmong patients’ cultural beliefs and values (e.g., shamanism), reducing fears by explaining common factors between patients' and providers' practices, and providing further information on the origin of cultural practices (i.e., herbal medicine). 

First, our providers emphasized the importance of communicating at practical levels by contextualizing their instructions for patients. During patients’ clinic visits, providers are required to deliver health information. In many cases, they are abstract, and far from patients’ everyday lives, as the Hmong patients’ lifestyles are often discrepant from those of the majority of Americans in the area. Providers emphasized the importance of transforming behavioral guidelines and healthcare information to align with Hmong patients’ behavioral norms based on their culture. During our interview, providers shared their approaches to capturing patients’ behaviors that could be easily changed or effectively supported by prescriptions, including understanding the cultural meanings of particular ingredients and the redefinition of ‘normal’ in shamanism. 

For Hmong community members, including many of the male patients, drinking alcohol with family and friends is considered a way of expressing respect, especially for individuals older than them. As a physician from the Hmong community, Provider 4 was well aware of the cultural meanings of alcohol consumption for his patients, which allowed him to deliver detailed guidelines that would help patients to have reduced amounts of drink. As a physician with a Hmong cultural background, he understood that the true meaning of "drinking" goes beyond simply consuming alcohol with others. It is instead listening to others while they drink in social contexts. After he captured the cultural meaning behind, he further explained the core meaning of this cultural practice to his patients. This led him to present a practical suggestion for his patients: recommending that they use his name as an excuse to avoid drinking, so as to reduce community pressure.

\begin{quote}
    \textit{“Some people [believe] that drinking shows respect, but it’s not necessarily the drinking part. It’s [about] the accepted part of all the words that are said by the other person. I just tell them that, ‘Hey, if you go to an event and they say, hey, drink this or drink that, just tell them that Dr. XX won’t let you.’ And so put the blame on me, because they have health issues and they really shouldn’t drink.” (P4)}
\end{quote}

As a registered nurse who has a Hmong cultural background, Provider 7 expressed her concerns about the increasing prevalence of diabetes among Hmong women. She found that when she advised her patients to reduce their consumption of highly processed, refined carbohydrate foods and those with added sugar, the concept was not immediately grasped. This was a significant communication challenge, which led her to come up with more practical approaches, using everyday tableware (e.g., a spoon and a cup) and physical references (e.g., objects to demonstrate the weight of ingredients), to reassure her patients that they could safely consume the main ingredients in their diet (rice and meat).

\begin{quote}
    \textit{“We don’t know if it’s [because of] stress [from taking care of the extended family members], if it’s family, if it’s the food that we eat, but there’s an increase in our Hmong population at the clinic of diabetes...[So, what] we gave to our Hmong women was, this is a spoon, this is one cup of rice, and this is what you should eat, and to help with trying to... we gave them cards, like what does three ounces of meat look like.” (P7)}
\end{quote}

Another effective communication strategy that providers have developed is the practice of showing understanding of Hmong patients’ cultural beliefs and values, such as their preferences for herbal medications and spiritual practices. During our interviews, providers emphasized the importance of showing understanding in providing comfort by accommodating patients’ values, accepting that traditional practices are not necessarily harmful to patients’ health. This approach makes both providers and patients feel acknowledged and appreciated during their clinic visits. 

Providers 12 and 10—one a physician with a Hmong cultural background and the other a White American social worker—both emphasized the importance of demonstrating an understanding of Hmong patients' cultural practices. Regardless of their own cultural backgrounds, both providers agreed that showing awareness of Hmong culture contributes to positive experiences for their Hmong patients. Provider 12 stated that many of his patients engaged in shamanic ceremonies (e.g., performing animal sacrifice, consuming specific food, and believing in spiritual figures) as integral parts of their culture. He does not seek to prevent them from following these traditional practices. Instead, he encourages a balance between these cultural practices and modern medical treatments. He believes that this balance can bring comfort and relief to his patients, which eventually leads to positive health outcomes.

\begin{quote}
    \textit{“So I get to hear that story and I’m trying to understand as well that this is a part of their belief and it’s also very helpful in their healing. So for me to be able to respect that and honor that without having to say you need to be on medications…The shamanism practice and stuff, I say, ‘You know what? If that helps you feel good, you meet your spiritual side, go for it.’ That’s like going to church and praying and whatever.” (P12)}
\end{quote}

Several providers shared communication strategies to improve their patients’ medication adherence directly. Providers 7 and 12 told of their practices to reduce patients’ unfamiliarity with Western medicine, including medical examinations and medication prescriptions. By explaining commonalities and relationships between patients’ and providers’ practices, providers were able to improve patients’ understanding of the care process. In the Hmong community, a wide range of herbal medicines, including brewed tea and creams, are commonly used to treat diseases or maintain health. Provider 12 shared that many medicines used in Western medicine are also made from plants that have been processed. For him, explaining these commonalities helps alleviate his patients’ fears, so they can improve their medication adherence.

\begin{quote}
    \textit{“And so it’s still very, I want to do herbal stuff, because that’s natural and that’s better for, that’s still kind of the mentality. And I chatted, [telling] them this medicine comes from herbals too, they come from herbs.” (P12)}
\end{quote}

Although they are favorable towards herbal medicines used in the Hmong community, providers pay extra attention and are cautious about practices that can cause side effects. This can be a problem if patients are using both their prescribed medication and their herbal medicine. To prevent these adverse outcomes, providers share extra information on herbal medicine, such as the possible side effects of a particular medicine and the potentially harmful impacts of taking medication and herbal remedies simultaneously. Provider 7’s patients often overtrust herbal medicines, which can cause side effects. She emphasized the necessity of knowing which herbs can be harmful and which are safe enough to consume in combination with their prescribed medications.

\begin{quote}
    \textit{“There's nothing wrong with it because they're roots and herbs and stuff like that that they can drink. Yes, but there's also a part where the Hmong community don't understand that these herbs can also have side effects mixed with your medication. They hear that in the community like, ‘Oh, my gosh, you take this [Western medication], and they die.’ But then they think it’s the doctor who gave them medication that they died from, not the Hmong herbs, but it's opposite sometimes.” (P7)}
\end{quote}

This section shows the benefits of providers acknowledging patients’ beliefs in cultural practices and possible fears and unfamiliarity with Western medicine during their interaction with patients. Instead of strictly imposing prescriptions and making patients comply, providers emphasized the importance of understanding patients’ cultural backgrounds, showing acknowledgment, and providing flexibility and extra education on Hmong patients’ practices.

\subsection{Strategy 2: Understanding the Value of Strong Social Ties and Collective Decision-making Practices in the Community}

Providers’ experiences showed that their approaches when communicating with Hmong patients can be significantly influenced by the patients' close community ties. Social relationships play a decisive role in health management and overall lifestyles throughout Hmong patients’ lives, including diet, health behaviors, and treatment plans. Hmong patients often value wisdom from elders, which influences their reliance on authoritative family members (e.g., male breadwinners) for their critical medical decisions.

The strong impact of solid ties in health management results from a unique social structure. The Hmong community consists of clans based on family names, and members of each clan share ancestors and form strong connections grounded in their kinship. Clan members commonly refer to each other in terms of their parents or grandparents. Hmong patients tend to consider the impacts of medical treatments on family dynamics or their family’s reputation (e.g., the stigma surrounding chronic illness) beyond their own lives.

Together, these characteristics in the Hmong community have influenced the unique ways in which Hmong patients communicate with their providers and manage their health. These strong social ties led providers to seek ways of using them during communication with Hmong patients, including actively involving a community or family leader during the decision-making process, sharing anecdotes from the community to resonate with patients, increasing relatedness using providers’ connection with the community, and constantly increasing understanding of unique cultural practices. 

For our providers, one effective way to facilitate patient communication is to engage a community or family leader—often the patient's father or husband. Engaging a family/community member who has social power often leads patients to be more responsive. Provider 15 shared the valuable practice of holding a care conference. These conferences are collaborative, and involve not just providers and patients, but also patients’ families and community members, which allows all parties to openly share their perspectives on treatment options and potential consequences. This helps in deciding on the best treatment plan. As a care coordinator with a Hmong cultural background, she shared her strategy of identifying trusted community leaders to help influence patients and their families in selecting the appropriate treatment options.

\begin{quote}
    \textit{“What I'm going to do, I find out who your leader is, who can convince you to make the decision, so someone in the family they will trust, someone in the Hmong culture they will lead the family. … So the leader will come into the family saying that I think it's a good way to have the surgery, and many times is resolved like that.” (P15)}
\end{quote}

Provider 7, a registered nurse, also mentioned her strategy of involving patients’ families, saying that the family is very important to Hmong patients and is an important part of their health management.

\begin{quote}
    \textit{“I always involve their family...I involve their family, because the family is very important to them and also an important part of their healing process and also for their health.” (P7)}
\end{quote}

Similarly, Provider 13 suggests that her patients discuss their health issues with people in their close network and encourage them to return to the clinic.

Next, providers highlighted the effectiveness of sharing personalized anecdotes from the Hmong community. They pointed out that these anecdotes are a valuable tool to inform about the potential consequences of their choices in a way that resonates with patients. Recounting actual cases that happened to Hmong community members is more beneficial than providing general health information, and leads patients to take providers' suggestions more seriously. Provider 12 mentioned how effective it was to remind patients of community members who developed severe symptoms due to non-adherence to medication.

\begin{quote}
    \textit{“And everybody knows everybody who has lost a feet or had a stroke in the community. I will explore with them. Oh, so you mentioned maybe your mom had diabetes. Okay. What happened to her? So using that to relate to them like, ‘Oh, this could happen to me too.’ And so I think for me, my challenge is finding something that's relatable to them, whether it be a cousin, a friend, a family, someone that they know that has been through this that will hopefully bring some sort of awareness.” (P12)}
\end{quote}

Provider 6, a White American physician who had no Hmong cultural background before working with Hmong patients, shared her strategy of describing similar cases from patients' families to foster understanding and build trust. She mentioned that general, non-personalized health information did not lead patients to follow prescriptions.

\begin{quote}
    \textit{“And so you don't want to say, ‘Hey, you have diabetes, if you don’t get your A1C under seven, you’re going to have a heart attack, you’re going to have a stroke, you’re going to die.’ [Instead,] you talk about sometimes they’ll bring it [the stories in the community] to you, ‘My brother had a stroke.’ ‘Well, did your brother have diabetes? Did your brother have high blood pressure?’ [So,] this is what we want to control, or talk about.” (P6)}
\end{quote}

Providers also shared their efforts to increase patients' relatedness by identifying and utilizing their connection with the Hmong community. For example, Provider 4 often asks patients questions about their communities and families to identify personal connections. 

\begin{quote}
    \textit{“You say hi, you ask, how's your family? You ask about other kids, their parents, and you jot down things about what they talked about last time.” (P4)}
\end{quote}

Similarly, Providers 7, 11, and 12 address Hmong patients as if they were providers’ family members. In the Hmong community, people often refer to each other as relatives. Although these providers were born and raised in the U.S., their Hmong cultural heritage made it more natural for them to address Hmong patients as extended family. Providers 11 and 12 refer to older patients as “auntie, grandpa, or uncle,” which not only shows respect for elders but also builds a strong sense of community. Provider 11 shows her genuine feelings and intentions for her patients by treating patients like her family, further reinforcing the trust in the patient-provider relationship.

\begin{quote}
    \textit{“And I call everybody auntie, or uncle, or grandma, or grandpa. So it's kinda like saying, ‘I am your sister and I am your kind and I'm here to help you,’...‘I'm just telling you this because I see you like you would be my mom and if it was my mom...” (P11)}
\end{quote}

Similarly, Provider 7 notes that her patients often address her as "daughter", even though they are not related by blood. This practice reflects the Hmong community's friendly and warm nature. 

\begin{quote}
    \textit{“They don't say, "Nurse," or anything. They're calling you, ‘Daughter, how does this help me? What is the doctor trying to say? Can you relay this to my doctor? And can you help me with this?’” (P7)}
\end{quote}

Also, providers constantly try to increase understanding of the unique cultural practices of the community. Providers indicated that working with Hmong families improved their general knowledge of the Hmong religion, cultural traditions, diet, and the role of family and community in decision-making, and specifically made them aware of patients’ situations and relationships with their community members. Provider 13, who recently started her career as a nurse practitioner, was born and raised in the U.S. and is, therefore, more familiar with American culture. She shared her learning experiences from fellow nurses who had a natural inclination toward Hmong culture. She emphasized the importance of gaining an in-depth understanding of the Hmong community, noting that such knowledge fosters trust and enhances communication with patients.

\begin{quote}
    \textit{“There are a couple of nurses who are able to have great relationships with these Hmong families, because they work a lot with these Hmong families. And so they know exactly what their culture entails, how family decision making is made, how Hmong families are like, for lack of better words…” (P13)}
\end{quote}

However, strong social ties are not always beneficial: the involvement of family and community also causes rumors and misinformation to spread quickly. For instance, providers shared that there are stories of Western medicine causing the deaths of community members, even though the patients were already very sick when they were brought to the hospital. These stories often stem from a lack of understanding about the nature of the illness and the limitations of medical treatment. This further leads to community members' disinclination to heed providers' prescriptions. This requires extra attention to mitigate the negative impact of the community.

This subsection shows the important role of the Hmong community's strong social ties, in providers' communication with Hmong patients. Strong social ties can be a powerful tool that effectively fosters communication between patients and providers. Providers suggested seeking the active involvement of a community or family leader during the decision-making process, sharing community stories to achieve greater resonance with, increasing relatedness via providers’ connection with the community, and making constant efforts to understand the unique cultural practices of the community, as approaches to maximize the benefits of the community.

\subsection{Strategy 3: Establishing Trustworthiness through Gradual and Inclusive Long-term Care}

Providers commonly pointed out that establishing trust with Hmong patients is essential to achieving effective healthcare outcomes. Hmong patients are particularly concerned and fearful when prescribed medications do not work immediately or when side effects cause discomfort. During the interview, providers shared Hmong patients' preference for having a trustworthy, long-term relationship with their healthcare providers, which is related to their status as war refugees. Because of their negative experiences when leaving Vietnam for the U.S., including substantial betrayal and prejudice, the Hmong are likely to interact with those from their close network. Furthermore, according to our providers, language barriers and cultural differences have historically led to Hmong patients being misdiagnosed with psychotic disorders. They experience frequent miscommunication and receive prescriptions without proper explanation of side effects, which has hindered the establishment of effective rapport with healthcare providers.

The negative experiences lead to reducing their recommended dosage—if they do not discontinue their medication entirely—and visiting clinics only rarely. Hmong patients, if not in a trusting relationship with their providers, often change their primary providers in search of better treatments; as a result, providers miss important contexts, make inconsistent treatments, and risk misdiagnosis.

To build long-term relationships and rapport with Hmong patients, providers developed helpful strategies to build trust with new patients. Specific strategies include gradually introducing new medications and therapies, offering a time frame to consider possible treatment options, communicating in a respectful manner, and frequently following up via phone calls. After initially developing a rapport with the Hmong patient, providers maintain and solidify the patient’s trust. Hmong patients start coming to providers for follow-up visits and prefer to maintain a long-term relationship.

When introducing Hmong patients to Western medications and therapies, providers take a gradual approach instead of addressing multiple problems at once. This strategy is intended to minimize drastic changes and alleviate reluctance when adjusting patients' treatment plans. Provider 9, who does not have Hmong cultural affiliations, shared her experiences to build trust with new Hmong clinic patients. 

\begin{quote}
\textit{“So it's the first time I'm seeing them, it's a little harder to kind of pinging their trust. Follow-up patients, usually they will have seen me or my colleagues form a few times. So usually the big difference is that they kind of understand your workflow, and they understand how they trust you, and they're willing to come back to see you again.” (P9)}
\end{quote}

Similarly, Provider 8, whose family has Hmong cultural origin, always encourages first-time visiting Hmong patients to follow healthcare practices that they are comfortable with, rather than strictly asking them to follow prescribed medications. According to her, it takes several visits to make them comfortable enough to trust providers' prescriptions.

\begin{quote}
    \textit{“It takes several visits for them to see, okay, maybe the first visit they don’t see that their condition is affecting them, but then the second visit, they come back maybe with more symptoms after talking about lifestyle modification. And then the next step will be, ‘Okay, well, we’ve tried, we worked on this already, let’s try this. Let’s just see if this is going to help improve the symptoms that you’re having.’ And having them try what they feel comfortable with first. And if it’s not working, then we can try what we are recommending.” (P8)}
\end{quote}

Regardless of their affiliation with the community, providers shared the importance of the gradual introduction of unfamiliar medical information and practices. As a White American who does not share similar cultural backgrounds with Hmong patients, Provider 6's primary goal is to minimize the confusion that can be caused by drastic changes, including multiple new medications. Based on her experiences, patients will stop adhering to a new medication if it causes a side effect, even if the side effect is minor. Similarly, Provider 15, who is familiar with both Hmong and American culture, shared how providers present new medications to Hmong patients. This “one step at a time” strategy is slow, as it takes several visits to settle into the new prescription, but it is crucial for building trust with patients and ensuring greater adherence.

\begin{quote}
    \textit{“So the doctor said then we start in zero. If you don’t trust me, I’m going to start one medicine with you first, and we will start with the second one, and work forward. And little by little, they build the trust between the doctor and the patient. And they start to trust each other, and the doctor want her to come and see every three months, the patient will go.” (P15)}
\end{quote}

Another useful strategy for building trust was setting a time frame for Hmong patients to consider their treatment options and reconnect with providers. Providers 5 and 15 mentioned that they often suggest a couple of hours to a few weeks so that patients do not feel pressured or forced to make immediate decisions on completely new treatment options.

\begin{quote}
    \textit{“And give a time, say I give a couple minute, a 30 minute or one hour, two hour for you to talk about your family and come out with the decision, or I don't have no time, but I tell you the pro and the con, and say yes or no.” (P15)}
\end{quote}

Provider 5 takes a similar approach, providing patients with a time frame to consider potential treatment plans. As a care coordinator, she also highlighted the value of checking patients’ status over the phone. She would call her patients in the following days to see if they had any questions about prescriptions, a practice that significantly increase the level of trust and led to positive care outcomes.

\begin{quote}
    \textit{“Most of the time there's just a lot of information. Then, I just, ‘Take it home. Let it sink in, and I'll connect with you in a couple of days and connect with you to see how you're doing and if you have any questions at that time.’” (P5)}
\end{quote}

Providers also shared respectful ways of communication as a key to developing trust with Hmong patients. Many providers pointed out that the Hmong community values interaction where they are cared for. Providers explained how they demonstrate their trustworthiness by showing respect for patients: explaining the logistics of their visits in addition to diagnoses in a detailed manner, which creates a safe environment for patients. Regardless of their cultural backgrounds, Provider 10 (a White American social worker) and Provider 15 (a care coordinator of Hmong origin) both emphasized the importance of Hmong patients perceiving that their healthcare providers take their health seriously. Provider 15 always greets patients and shares detailed information on their visits, including their upcoming schedule. This level of detail gives patients the feeling that they matter to their providers. When she showed her genuine care through these consistent acts, patients trusted her and considered her a good healthcare provider.

\begin{quote}
    \textit{“I always greet them, they say hi, in the morning, tell them who I am. This is this from this. … when they're done with the visit, I always say bye, see you next time. Don't forget our next visit will be what time, and I will remind you when the day come so you know that we are going to have that appointment. … So most people love that, the interaction, the welcoming, introduce myself who I am, that say goodbye and say repeat and call them to remind them for next appointment...So when those consistency, it's built a lot of trust to [the Hmong] community.” (P15)}
\end{quote}

Providers also frequently follow up with Hmong patients through phone calls to show their care for patients, which influences long-term, trusting relationships. This frequent communication not only ensures the patients' rapport with their providers but also significantly enhances patients' health outcomes. For example, care coordinators talk weekly on the phone with Hmong patients who have chronic conditions, including diabetes, and advise the patient to follow guidelines (e.g., insulin) and track their records to report back to them. Provider 3, a White American physician, shared her patient's positive experience with frequent communication through their care coordinator (a liaison between patients and physicians).

\begin{quote}
    \textit{“And especially if you have kidney disease, you can't use certain diabetes pills. And so she was just really stuck between a rock and a hard place. But her care coordinator has been really good, calls her once a week, elicits her concerns, tells me her concerns, then calls her back. And so she's just constant communication with her to try. I mean, it is working. She's on insulin.” (P3)}
\end{quote}

When patients establish trust with their providers and start to feel safe, they engage in long-term care, keep seeing the same provider for many years, and refer their family members to that provider. These referrals are the signs of patients’ strong trust, making healthcare providers feel more valued and appreciated. Regardless of their cultural affiliation with the Hmong community, providers shared that when their patients see them for many years and bring their family and friends to them, they realize that they have earned their patients' trust. Provider 14 mentioned that long-term relationships make providers feel positive about their work and the community.

\begin{quote}
    \textit{“But we still had the patient who'd been coming for many years still come. … So I think they're happy and I'm happy to see them. Sometimes we see each other and when I see the kid grow up, they say, ‘Oh, remember me, I'm this person.’ I say, ‘Oh, my goodness, you grow up.’ They were baby, they changed.” (P14)}
\end{quote}

This subsection shows the importance of building trust between healthcare providers and Hmong patients to deliver effective healthcare for the community. Providers use different strategies, such as gradual medication introduction, a time frame for considering possible treatment options, respectful communication, and frequent phone follow-up. Once patients trust their providers, these relationships turn out to be long-term ones that eventually become the most effective and positive practices for both patients and providers.

\subsection{Strategy 4: Adapting Communication Style Based on the Particular Characteristics of the Hmong Language}

Our interviews with healthcare providers, particularly physicians and nurses, highlighted the limitations of the Hmong language in delivering health information. The Hmong language lacks many medical terms, such as the names of medicines and symptoms and the concept of preventive care. This linguistic gap makes it difficult for providers to communicate effectively with patients. Furthermore, due to the relatively small population of Hmong speakers, there is only a small number of health education resources designed for Hmong patients, including discharge documents. Also, the indirect nature of the Hmong language presents additional barriers, as direct expressions can often frighten patients. 

Providers have developed effective communication strategies with Hmong patients, drawing on their experiences to ensure patients understand their care. They use storytelling to describe symptoms, capture patients’ non-verbal cues, adjust their tone when discussing diagnoses, and use visual materials (e.g., pictures and videos) to explain pathology, medications, and diet.

First, during the interview, participants shared their strategies for using storytelling as a communication approach. They encourage Hmong patients to share their symptoms by asking them to use storytelling, allowing them to spend more time describing their symptoms and providing contextual information. Providers also use storytelling approaches when explaining medical terms, in particular body parts and names of diseases, that are unavailable in Hmong. Traditionally, the Hmong community has passed down their history, wisdom, folklore, and knowledge through their storytelling tradition. Until the 19th century, the Hmong community relied heavily on oral tradition, which is still the primary mode of communication for some Hmong community members. As a physician who has a Hmong cultural background, Provider 12 mentioned that he allows his patients to share stories, even if it means exceeding their scheduled appointment time, because, as someone from the community, he understands the importance of oral communication for his patients. He found it valuable to listen to his patients’ stories, as they often included critical health information that might not otherwise have been mentioned.

\begin{quote}
    \textit{“I get a case, but I think for my older population, I let them tell their story, because they like to tell their story and it can take a 15-minute appointment becomes like a 45-minute appointment, but at the last five, ten minutes, that's when I get the juicy stuff, things that I need to know.” (P12)}
\end{quote}

As a clinical coordinator and an interpreter, Provider 15 shared her experiences of explaining English medical terms with storytelling. It is particularly useful for explaining to her patients about the intricacies of the American healthcare system, which is impossible in Hmong. Similarly, Provider 7, said that she has difficulties in explaining certain anatomical details in Hmong, and has only been able to make patients comprehend by telling them long stories. Although she is affiliated with the Hmong community and relatively fluent in Hmong, she noted that explaining medical terms in Hmong requires additional effort. She shared that it is not simply a matter of translating words, but involves conveying complex concepts in a culturally meaningful way.  

\begin{quote}
    \textit{“There are certain body parts that we have that are not... we can't really translate that into Hmong. Some wordings that we have doesn't translate into Hmong, and so sometimes we have to tell a long story to have the patient understand the process.” (P7)}
\end{quote}

The importance of storytelling in the Hmong community was emphasized by both groups of providers—those with Hmong cultural heritage and those without. Provider 6 shared that after more than 20 years of working with Hmong patients, she has come to see herself as a storyteller.

\begin{quote}
    \textit{“As a provider from a middle-class, white background, I connected with my patients and have made this my career…I have changed a lot. I have become much more of a storyteller, which is very much more the way it was. I'm not sure my patients haven't changed me more than I've changed them at all. (P6)}
\end{quote}

As many Hmong patients speak little to no English, providers pay extra attention to patients’ nonverbal communication cues to improve their understanding of patients' symptoms and conditions, such as gestures and facial expressions. Capturing these nonverbal cues assists providers in recognizing potential health-related problems that patients and their caregivers easily overlook or cannot articulate. For Provider 5, catching subtle nuances through patients’ gestures allows her to focus on their needs and ask follow-up questions.

\begin{quote}
    \textit{“Communication goes, I think just understanding the nuances even in the shift in body language for them or just certain words that they say you pick up and you’re like, ‘What's going on?’” (P5)}
\end{quote}

Similarly, as a nurse practitioner, Provider 13 believes observation is a huge part of her job, as it allows her to capture patients' and families’ feelings while they speak.

\begin{quote}
    \textit{“So it's our job to assess and observe. Observation is such a huge part of our job. And so even if families don’t tell you how they’re feeling, your observation tells a lot about how they're feeling, and what these families need.” (P13)}
\end{quote}

Another effective strategy providers have shared is the use of a less direct tone when discussing patients' diagnoses and the potential consequences. By choosing their words carefully and using culturally appropriate expressions, providers can significantly reduce patients' concerns and fears. Provider 14 is an interpreter who assists the patients during their clinic visits. She shared an example of translating a diagnosis in a less direct form of expression.

\begin{quote}
    \textit{“‘Your kidney is not working right.’ And then you have to find a way to say so they won't be panicked or they won't be worried. Yeah. [the physician] can say, ‘Oh, your kidney is fail.’ And I say it's not working good or it's not working as good as supposed to. I explain like that. If you say your kidney failed, they're probably scared. I kind of use the word that motivate them to say, ‘Oh, okay. That means it's not working as good as supposed to?’ But you have to find a way to explain so they won't be worried too much or scary.” (P14)}
\end{quote}

As a clinical interpreter, she helps physicians to avoid inappropriate phrasing in Hmong, and use more indirect ways to deliver health information. This allows patients to interact with providers without unnecessary fear.  

Lastly, regardless of their cultural background, providers frequently use visually oriented communication strategies when explaining pathology and recommending specific lifestyle changes to Hmong patients. For example, to illustrate the causes of illness (e.g., Hepatitis B) and associated symptoms, Provider 3, who is a White American physician, often uses pictures of the liver and viruses.

\begin{quote}
    \textit{“I'm trying to explain hepatitis B. I'll have a picture of the liver and a little picture of a virus and a picture of yellow eyes to explain the symptoms of Hepatitis B would be an example.” (P3)}
\end{quote}

In addition to illustrating symptoms and pathology, visual materials are also effective resources to guide patients in learning about necessary lifestyle changes. Provider 8 uses images from the CDC (Center for Disease Control and Prevention), showing them to patients as references. Provider 6 also actively uses Google Images and YouTube videos to advise patients to exercise regularly. For her, those multimedia materials are the biggest patient education tool, as many of her patients have language barriers. When she explains numbers or particular types of medications, she always shows patients images. Other healthcare providers, including Provider 7, who has Hmong cultural heritage, use illustrations to enhance patients' understanding. She has found that drawing the moon and the sun to indicate the right time for taking medication, rather than just stating the time, is an effective communication strategy. This use of simple images has reassured her that patients effectively receive necessary health information during their clinic visits. 

\begin{quote}
    \textit{“And you draw the sun, you draw the moon, morning appointment, morning meds, afternoon meds, or the pill box. You separate them and you have their family involved.” (P7)}
\end{quote}

This section shows providers’ strategies to address barriers associated with Hmong patients’ health literacy and the particular characteristics of the Hmong language. They use storytelling to describe symptoms, capture patients’ non-verbal cues, and adjust their tone when discussing diagnoses, which helps alleviate fear in Hmong patients. They also use visually oriented materials, like pictures and videos, to explain details. These strategies enhanced Hmong patients’ understanding of providers’ explanations of diagnosis and symptoms during clinic visits.

\section{DISCUSSION}

Addressing RQ1, our study identified four types of communication strategies that not only enhance providers’ diagnosis, treatment, and care during patients’ clinic visits, but also facilitate interpersonal relationships between healthcare providers and patients. These strategies, when implemented, have the potential to create a positive and sustainable healthcare experience for Hmong patients and their families, thereby improving their long-term health management. Providers considered that patients’ rationales for clinic visits went beyond effective treatment and fast recovery. Patients’ compliance with prescribed treatment is significantly influenced by their perception of providers, in terms of general trustworthiness, and particularly in terms of how they communicate diagnosis and instructions. In our study, providers recognized their roles and responsibilities in alleviating the Hmong community's unfamiliarity and discomfort with the healthcare system in a developed world, which stems from the community's history as refugees. In this section, we address RQ2 and RQ3 by discussing the multiple roles of cultural competence in patient-provider communication in marginalized communities, and proposing design implications for technology that could facilitate providers’ cultural competence and support providers to communicate with patients from these communities.

\subsection{Cultural Competence for Facilitating Patient-Provider Communication in Socially Marginalized Communities}

Cultural competence is a set of attitudes and behaviors that enable individuals to work effectively in a cross-cultural environment~\cite{anderson_etal_CulturallyCompetentHealthcare_2003, lau_rodgers_CulturalCompetenceRefugee_2021}. This concept has been widely used in different domains, including public health~\cite{lau_rodgers_CulturalCompetenceRefugee_2021}. Previous studies showed how cultural competence is applied in various contexts, including medical (e.g., mental health~\cite{rice_harris_IssuesCulturalCompetence_2021}, rehabilitation~\cite{grandpierre_etal_BarriersFacilitatorsCultural_2018}, nursing~\cite{loftin_etal_MeasuresCulturalCompetence_2013}) and non-medical fields (e.g., education~\cite{eden_etal_CulturalCompetenceEducation_2024}). For example, researchers have developed training programs that empower healthcare providers to enhance their cultural competence. They have highlighted the importance of involving providers in tailored education programs, interpreter services, and the recruitment of bicultural staff~\cite{beach_etal_CulturalCompetenceSystematic_2005, hollifield_etal_EffectiveScreeningEmotional_2016, lau_rodgers_CulturalCompetenceRefugee_2021}. When cultural competence is integrated into healthcare settings, it improves understanding among stakeholders involved in the care, access, and effectiveness of healthcare services for patients of diverse ethnicities, values, and behaviors~\cite{anderson_etal_CulturallyCompetentHealthcare_2003, lau_rodgers_CulturalCompetenceRefugee_2021}. Studies found that when providers have increased cultural competence through cultural awareness, cultural knowledge, and cultural skills/behaviors of patients~\cite{alizadeh_chavan_CulturalCompetenceDimensions_2016, shen_CulturalCompetenceModels_2015}, patients have positive care experiences. These influences of biases and communication ability have been actively studied to explain the importance of cultural competence~\cite{alizadeh_chavan_CulturalCompetenceDimensions_2016}. 

During our interviews, the providers mentioned that they developed cultural competence from their experiences with the immigrant patients, and how it was applied to their practices in routine care. For example, providers explained how they showed their understanding of Hmong patients’ cultural beliefs and values, such as the origin of herbal medicine and shamanism, and reduced fears by showing commonalities between patients’ and providers’ practices. They applied their cultural knowledge to their communication with patients. 

Given that most of our participants have varying levels of affiliation with the Hmong community, we recommend facilitating the involvement of healthcare providers who share a similar cultural background to enhance cultural competence in communication when serving similar immigrant communities. Although we did not specifically focus on providers' positionality and its impact on communication with patients, several providers naturally highlighted the advantages of their cultural ties during the interviews. They shared insights into how being part of the community helps them better understand patients' experiences—such as utilizing their mutual connection with other community members, the cultural significance of drinking, the origins of herbal medicine, and the importance of identifying community and family leaders. While these strategies were shared by providers from other communities and can be valuable for healthcare providers who do not share the same cultural background as their patients, they may be particularly beneficial for those who do. By considering healthcare providers' unique positionality, including cultural background, there could be multiple technology-mediated approaches and opportunities to support providers' cultural competence. However, cultural competence has yet to be actively discussed in health technology design, with the exception of a few training tools designed for medical and nursing education~\cite{sakpal_VirtualPatientsTeach_2012}. The practical application of cultural competence to improve health management in immigrant communities has not been actively explored. Although prior studies in HCI examined approaches to improving health care in socially marginalized populations, the concept has not been applied to the design of health technologies in the field.
 
We suggest that HCI researchers incorporate the concept of cultural competence in health technology designed for immigrant communities. In this sense, the following subsection addresses RQ2 and RQ3 by discussing approaches to conceptualize cultural competence in health technology design and facilitate patient-provider communication for immigrant communities.

\subsection{Design Opportunities for Healthcare Communication Technologies in Immigrant Communities}

Building on previous studies on design considerations for health communication technologies in socially marginalized communities, we suggest two significant design implications for culturally competent health communication technologies.

\subsubsection{Culturally competent technology design through adaptive communication styles}

In our study, providers shared their approaches to adapting communication styles based on immigrant communities' unique characteristics and cultural practices. Their active use of different communication styles throughout the care process resulted in improved communication with their patients. In communication studies, communication style is the set of ways in which individuals send verbal or nonverbal signals during social interactions to indicate who they are, how they prefer to interact, and how their messages are interpreted~\cite{devries_etal_CommunicationStylesInventory_2013, norton_FOUNDATIONCOMMUNICATORSTYLE_1978}. One example of using different communication styles is adjusting the tone, such as the level of assertiveness (the degree of freedom in making choices) or the directness of word use (how explicit or straightforward the language is, compared to more indirect forms of communication influenced by sociocultural norms). Appropriate tone is a key factor in how effective a given communication style is in a specific social context~\cite{grinstein_kronrod_DoesSparingRod_2016}. 

In patient-provider communication, \textbf{providers’ tone, which includes their word choice, is a critical determinant of an improved communication experience}~\cite{buller_buller_PhysiciansCommunicationStyle_1987, he_etal_WarmCommunicationStyle_2018}. For example, one study showed the positive effects of providers using a more assertive tone when patients were experiencing more severe symptoms~\cite{kronrod_etal_WhenNeedsDont_2024}. Another study showed positive relationships between providers’ word choice (i.e., empathetic and less anxiety-inducing expressions) and patients’ satisfaction with their healthcare service~\cite{hall_etal_CommunicationAffectPatient_1981}. Other characteristics, such as friendliness~\cite{norton_FOUNDATIONCOMMUNICATORSTYLE_1978}, respect, and willingness to listen~\cite{devries_etal_CommunicationStylesInventory_2013} are types of communication styles that influence patient satisfaction~\cite{capone_PatientCommunicationSelfefficacy_2016}. 

In our study, providers shared their strategies for actively adjusting communication styles: they adjusted the language using a different tone (e.g., less direct words), explained complex medical information as a story to minimize patients’ concerns, and facilitated their understanding of the medical information. These strategies were implemented based on the premise that indirect communication styles are considered polite and appropriate in Hmong culture, whereas straightforward and clear communication are well regarded in the U.S.~\cite{kim_wilson_CrossculturalComparisonImplicit_1994}. The providers tried to develop cultural sensitivity and an in-depth understanding of their patients' cultural characteristics. They applied their strategies in interacting with patients, making them feel respected and valued. 

The strategies we captured from the interviews are well aligned with prior studies in HCI and related fields that showed improved satisfaction among service recipients when providers’ communication styles are congruent with those of the user group~\cite{seo_etal_LearningHealthcareProviders_2021, ghazali_etal_InfluenceSocialCues_2017, calisto_etal_AssertivenessbasedAgentCommunication_2023, rau_etal_EffectsCommunicationStyle_2009, rheu_etal_SystematicReviewTrustBuilding_2021}. When designing information delivery and communication technology to support socially marginalized groups, it is important to take into consideration their cultural characteristics. In the case of the Hmong community, \textbf{the technology should not explain symptoms in an overly blunt and straightforward manner nor jump directly to possible consequences.} Such an approach could easily discourage the patients’ active involvement in their care, reduce their trust in their providers, and lead to non-adherence to the treatment plans.

However, there is little study into the ways in which technologies could be designed to meet immigrant communities’ unique needs in communication styles~\cite{stowell_etal_DesigningEvaluatingMHealth_2018}. Existing research on immigrant communities focuses primarily on addressing their limited health literacy by delivering visually-oriented content~\cite{bender_etal_DesigningCulturallyAppropriate_2016, claisse_etal_UnderstandingAntenatalCare_2024} or language proficiency by translating medical terminologies~\cite{kim_etal_ItsMyLanguage_2024, thonon_etal_ElectronicToolsBridge_2021}. Effective communication styles, such as culturally accepted word choice and culturally appropriate ways of delivering content (in the Hmong community, storytelling), have not been studied well.

We suggest the active implementation of identified communication styles into the health communication technologies (e.g., patient portals). For example, when delivering health information for Hmong patients, such as after a visit summary and behavioral guidelines, \textbf{the system could present a set of appropriate stories in culturally well-accepted tones.} These platforms can be designed to be highly contextualized to culturally marginalized communities, taking into account each patient's background, age, and overall health condition. When providers share health information with the patient, these culturally competent platforms can suggest appropriate words, a specific set of stories and guidelines. If needed, providers can also apply content during their interaction. These stories and guides can be presented in personalized formats, such as video-, picture-, or audio-based materials, depending on patients’ backgrounds and preferences. Additionally, paper copies of text- or image-based digital materials can be sent to patients’ residences. For providers, this platform can show appropriate terms they can use, effective styles (e.g., direct vs. indirect expressions), and suggest contextual anecdotes and cases specific to the patient’s culture, all in a highly personalized manner. The findings on adaptive communication styles could be further applied to similar contexts, such as other racial and cultural groups. In particular, other immigrant communities or groups with LEP (e.g., children) could benefit from culturally appropriate, storytelling-based patient-provider communication technologies, which may help reduce health disparities in these populations.

\subsubsection{Culturally competent technology design that considers unique social dynamics in communities}

In line with studies on collaborative health management in close social networks~\cite{binda_etal_SupportingEffectiveSharing_2018, cha_etal_TransitioningIndependenceEnhancing_2022, schaefbauer_etal_SnackBuddySupporting_2015, silva_etal_CodesigningSituatedDisplays_2024, shin_holtz_BetterTransitionsChildren_2019, maestre_etal_NotAnotherMedication_2021}, the providers in our study shared the importance of understanding Hmong patients’ close social ties in communicating health information. They not only \textbf{invite patients’ family members and community leaders to critical health decision-making} but also \textbf{find connections with patients.} They also use connections with community members to communicate with patients, presenting anecdotes that resonate and fostering empathy. These strategies can directly be applied to technologies for collaborative health management, which utilize close social ties. 

Previous studies suggested technologies for collaborative health management in different social contexts, including families~\cite{shinDesigningTechnologiesSupport2021, shin_etal_MoreBedtimeBedroom_2022, shin_etal_BedtimePalsDeployment_2023, yamashita_etal_HowInformationSharing_2018} and broader communities~\cite{farao_DigitalStorytellingSupport_2023, harrington_etal_EngagingLowIncomeAfrican_2019}. For example, one study suggested a self-care tool for parents and children with diabetes. As children mature and become independent, their relationships with parents also change; the study suggested technology design opportunities for a shared understanding of managing diabetes in families~\cite{shin_holtz_BetterTransitionsChildren_2019}. Schaefbauer et al. proposed a mobile application for helping parents and children in low-income communities to monitor each other's snack intake, highlighting the usefulness of internal competition between two parties~\cite{schaefbauer_etal_SnackBuddySupporting_2015}. These studies showed that \textbf{when technologies are carefully designed depending on social dynamics and unique aspects of their relationships,} they can facilitate health information exchange, as well as related goals, such as better relationships and communication experiences~\cite{shin_etal_MoreBedtimeBedroom_2022, shinDesigningTechnologiesSupport2021, smriti_etal_MotivationalInterviewingConversational_2022}. In our study, because of the unique community-and family-based culture and dynamics in the immigrant community, when providers effectively utilize close social ties, they can easily improve patients’ adherence to the treatment plans.  

Our results also echo findings from previous HCI studies that emphasize social capital, such as strong bonds and close networks in marginalized communities, including immigrants~\cite{liaqat_etal_ExploringCollaborativeCulture_2023}, residents of the Global South~\cite{arueyingho_etal_AfrocentricCollaborativeCare_2023}, people of color~\cite{hui_etal_CommunityTechWorkers_2023, martin-hammond_purnell_BridgingCommunityHistory_2022}, patients and family caregivers in chronic illness~\cite{shin_etal_EveryCloudHas_2021}, and LGBTQ+ populations~\cite{li_etal_WeCriedEach_2023}. In these communities, members are closely tied and influence each other’s different aspects of lives, including health management~\cite{murnane_etal_PersonalInformaticsInterpersonal_2018, poortinga_CommunityResilienceHealth_2012}. One design implication is \textbf{a tool that allows patients to communicate key factors and values related to their health management, particularly those shaped by their cultural background,} such as unique cultural practices, important relationships within the community (e.g., identifying key family and community members involved in critical health decisions), boundaries for information disclosure, and the types of medical information they are comfortable sharing with family and community members. Previous studies showed that providers use online questionnaires, cameras, and wearable technology to collect information about patients' personal values, which are perceived to be less directly relevant to their symptoms. These tools also help coordinate the scale of information disclosure with their patients~\cite{berry_etal_SupportingCommunicationValues_2019}.

As our findings show, cultural values and their unique social dynamics in immigrant communities can significantly influence patients’ practices. For example, during the patient’s initial clinic visit, patients and providers (i.e., social workers and care managers) can work together to create a digital map rather than simply talking through all related information. The patient’s map could be used as guidance for providers to identify key personnel in their close network and explain the potential consequences of their health-related choices (e.g., explaining how their health decisions significantly influence their family and community). Such a design approach can enhance providers’ understanding of patients’ social relationships, which in turn can improve providers’ empathy in communicating with the patients.

\section{LIMITATION AND FUTURE WORK}

While we made efforts to recruit healthcare providers working with members of the Hmong community regardless of their cultural background, most of the providers who agreed to participate were individuals affiliated with the community. 10 out of 15 participants reported being fluent in both Hmong and English, due to their ties with the community (e.g., second- or third-generation members). One participant (P1) shared that, although he lacks fluency in the Hmong language, his family’s roots in the Hmong community allowed him to understand its culture. This indicates that these providers are highly committed to and motivated by their contribution to the community, which could play a key role in facilitating patient-provider communication. While we briefly mentioned providers' experiences with the community to communicate context in the results, we did not specifically explore their positionality during the interviews or how their personal backgrounds influenced their communication with patients in detail. Nonetheless, several providers expressed feelings of satisfaction and fulfillment from giving back to their Hmong community through their medical expertise. For providers without direct experience working with immigrant communities, the strategies identified in this study may not be immediately applicable. However, many of these strategies were also shared by providers who were not affiliated with the Hmong community, making them valuable for a broader range of healthcare providers. To gain a deeper understanding of how healthcare providers' lived experiences—including their cultural backgrounds—influence their communication with immigrant patients, future research could include a broader range of participants, particularly those from different cultural backgrounds, and compare their communication strategies to the findings of this study.

We also recognize that Hmong patients are not a homogeneous group with identical sociocultural backgrounds. Patients’ diverse identities—such as gender, socioeconomic status, immigration status (e.g., first-generation versus third-generation immigrants), and health conditions—can significantly influence their communication experiences with healthcare providers. For instance, a female patient with low literacy may struggle to effectively discuss gynecological issues with a male provider from the same culture due to concerns about privacy and the lack of adequate language skills. Although our interview protocols did not include questions on patients' intersectionality, given the varied experiences within immigrant communities, an intersectional framework could offer a valuable lens for exploring how multiple identities shape individuals’ unique healthcare experiences. In future studies, this approach can also be applied to understanding the health communication challenges faced by other marginalized populations, including those with chronic conditions, as well as individuals from the Global South and LGBTQ+ communities.

Additionally, the current study primarily focused on healthcare providers' perspectives on working with immigrant community members, rather than examining patients' communication experiences with their providers. To achieve a more balanced and comprehensive understanding of patient-provider communication within immigrant communities, future research should explore how Hmong patients perceive interactions with their providers and manage their health in the U.S. This would also help identify important design implications for healthcare technology, which could be investigated through suitable research methods based on the unique cultural characteristics of the community, such as co-design approaches (e.g., storytelling-based workshops).

\section{CONCLUSION}

In this study, we interviewed 15 healthcare providers who work with Hmong patients, and identified four different communication strategies. Previous studies examined factors that facilitate communication between healthcare providers and patients in different socially marginalized communities, and suggested technology-mediated interventions for alleviating challenges. However, they focused on the accessibility of healthcare services and health literacy mainly in developing countries. Also, only a few studies have discussed healthcare providers’ perspectives on communicating with patients from socially marginalized communities. Through interviews with our participants, we characterized four types of patient-provider communication in immigrant communities, and suggested design implications for health technologies for immigrant communities. We discussed how technologies could be designed to support cultural competence in patient-provider communication more practically. Identified implications can further inform opportunities for the design of supportive patient-provider communication technologies in broader culturally marginalized communities.

\begin{acks}
This work was supported by the University of Minnesota’s Social Justice Impact Grant. We would like to thank the healthcare providers who participated in the interviews, as well as the anonymous reviewers whose time and efforts substantially improved this paper.
\end{acks}

\bibliographystyle{ACM-Reference-Format}
\bibliography{sample-base}


\end{document}